\documentclass[a4paper,11pt]{article}
\usepackage{pos}
\usepackage[english]{babel}
\usepackage[nottoc]{tocbibind}
\usepackage{xspace}
\usepackage{arydshln}
\usepackage[export]{adjustbox}
\usepackage{floatrow}
\usepackage{url}
\usepackage{natbib}
\newfloatcommand{capbtabbox}{table}[][\FBwidth]

\newcommand{\hess}{H.E.S.S.\xspace}
\newcommand{\fermilat}{\emph{Fermi}-LAT\xspace}
\newcommand{\source}{HESS~J1813$-$178\xspace}
\newcommand{\psr}{PSR~J1813$-$1749\xspace}
\newcommand{\fermi}{4FGL~J1813$-$1737e\xspace}           

\title{Joint H.E.S.S. and Fermi-LAT analysis of the region around PSR J1813-1749}
 \ShortTitle{Reanalysis of the region around \psr}

\author*[a]{T. Wach}
\author[a]{A. M. W. Mitchell}
\author[a]{V. Joshi}
\author[a]{S. Funk}
\author{on behalf of the H.E.S.S. Collaboration}

\affiliation[a]{Friedrich-Alexander-Universität Erlangen-Nürnberg, Erlangen Centre for Astroparticle Physics,\\
  Nikolaus-Fiebiger-Straße 2, D 91058 Erlangen, Germany}

\emailAdd{tina.wach@fau.de}

\abstract{HESS J1813-178 is one of the brightest sources detected during the first HESS Galactic Plane survey. The compact source, also detected by MAGIC, is believed to be a pulsar wind nebula powered by one of the most powerful pulsars known in the Galaxy, PSR J1813-1749 with a spin-down luminosity of $\dot{\mathrm{E}} = 5.6 \cdot 10^{37}\,\mathrm{erg}\,\mathrm{s}^{-1}$. With its extreme physical properties, as well as the pulsar's young age of 5.6\,kyrs, the $\gamma$-rays detected in this region allow us to study the evolution of a highly atypical system. Previous studies of the region in the GeV energy range show emission extended beyond the size of the compact H.E.S.S. source. Using the archival H.E.S.S. data with improved background methods, we perform a detailed morphological and spectral analysis of the region. Additionally to the compact, bright emission component, we find significantly extended emission, whose position is coincident with HESS\,J1813-178. We reanalyse the region in GeV and derive a joint-model in order to find a continuous description of the emission in the region from GeV to TeV. Using the results derived in this analysis, as well as X-ray and radio data of the region, we perform multi-wavelength spectral modeling. Possible hadronic or leptonic origins of the $\gamma$-ray emission are investigated, and the diffusion parameters necessary to explain the extended emission are examined.}

\ConferenceLogo{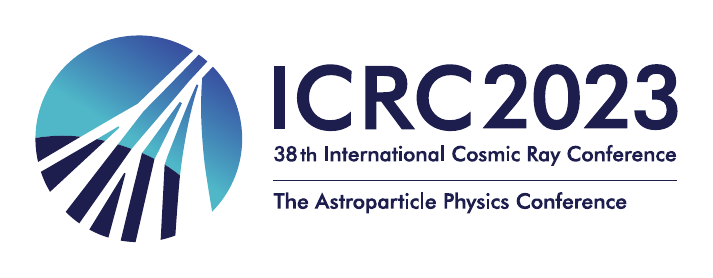}

\FullConference{%
38th International Cosmic Ray Conference (ICRC2023)\\
  26 July - 3 August, 2023\\
  Nagoya, Japan}

\begin{document}
\maketitle

\section{Introduction}
The Imaging Atmospheric Cherenkov Telescope (IACT) Array H.E.S.S. has highly contributed to our knowledge of the $\gamma$-ray sky, by identifying many previously unknown sources during their surveys of the galactic plane (HGPS) conducted in 2006 \cite{HGPS} and 2018 \cite{GPS}. One of these sources is \source, a single-component compact source, centered at $\text{R.A.}=(273.40\pm 0.005)^\circ$, $\text{Dec}=-(17.84\pm 0.005)^\circ$  with an extension of $\sigma = (0.036 \pm 0.006)^\circ$. Compact counterparts to this high energy $\gamma$-ray emission have been observed in X-ray with INTEGRAL \cite{integral} and XMM-Newton \cite{xmm-newton} and in radio with the VLA \cite{VLA}. Additionally, faint extended emission enclosing the bright compact emission has been detected during the HGPS in 2006. This detection however could not be established as significant \cite{HGPS}.

The detection of the shell-type Supernova Remnant (SNR) G12.82–0.02 \cite{VLA}, and the very young and energetic pulsar \psr, with a characteristic age of $5600\,$years and a spin-down luminosity of $\dot{\text{E}} = 5.6 \times 10^{37}\,\text{erg}\text{s}^{-1}$ \cite{psr}, both positionally coincident with the detected TeV emission, result in two different scenarios for the origin of the $\gamma$-ray emission. Several studies showed that it is more likely that the compact $\gamma$-ray emission observed in the TeV range is a result of Inverse Compton (IC) scattering of electrons from the pulsar with photons of the ISM \cite{xmm-newton,lepton_model2}

An analysis of the region in the GeV energy range, utilising data taken with the \fermilat satellite, showed emission that is positionally coincident with \psr, but extended $(0.6 \pm 0.06)^\circ$ \cite{fermi_araya}. They concluded that the origin of the TeV and GeV emission should be considered separately, and found that the origin of the $\gamma$-ray emission in the GeV energy range is most likely Cosmic Rays (CRs) accelerated at the shock fronts of the SNR. An additional hadronic scenario could be CRs accelerated at the shock fronts of the young stellar cluster Cl~J1813$-$178, located in close proximity to the pulsar as projected along the line of sight \cite{stellar_cluster}. Due to the mismatch in observed emission extension, and the very different interpretation of the data acquired in different energies, the origin of the emission in the region around \psr could not be firmly established yet. 

We present a reanalysis of the TeV data in the region around \source, using improved background rejection and event reconstruction. We also present a new analysis of the \fermilat data, with increased exposure compared to the previous work, and perform a joint-likelihood minimisation of the GeV and TeV data simultaneously. 

\section{H.E.S.S. Data Analysis and Results}
\hess data is taken in $\sim 28\,$min observation intervals referred to as runs. We select the runs for this analysis based on requiring that all four telescopes must participate, and the pointing position of the telescopes, which must be less than $2^\circ$ offset from the position of \source derived in the previous analysis. Additionally, standard data quality cuts \cite{sys_index} are applied and the energy threshold of the analysis is set to $0.4\,$TeV. Applying these quality cuts results in a dataset with $31\,$hours of livetime, taken between 2004 and 2012. We then analyse the data using the open-source Python package \texttt{gammapy, version 0.18.2} \cite{gammapy}. 

The first galactic plane survey \cite{HGPS} was evaluated using the Ring background method. This method estimates the background from source-free regions at equal distance from the source. The results acquired from this background method depend on a correct estimation of the source extension and is highly sensitive to point-like sources, but not suitable for faint or very extended sources and problematic for sources without a defined edge, like pulsar wind nebulae. 
In order to increase the sensitivity towards large extended sources, we use a template-based approach to describe the background. This template is generated from a large set of source-free regions, following the scheme in \cite{fov-bkgmodel}.

\begin{figure}
\centering
\includegraphics[width=0.8\textwidth]{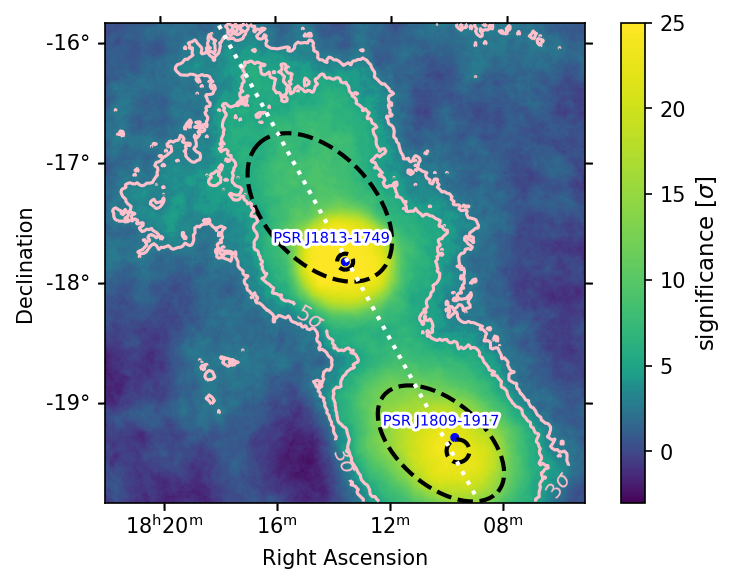}
\caption{Significance map of the region around \psr as seen by \hess before removal of emission using source models. The galactic plane is indicated by the white dotted line, $3\,\sigma$ and $5\,\sigma$ contours are shown in pink. The best-fit source model for HESS~J1809$-$193 derived in \cite{J1809} is shown by the black dashed line and the best-fit model for the emission around \psr derived in this study.}
\label{hess_sig}
\end{figure}

Figure \ref{hess_sig} shows the significance map of the region around \source with a correlation radius of $0.4^\circ$. 
The position of the galactic plane is indicated by the white dotted line. We also show $3\,\sigma$ and $5\,\sigma$ contours of the emission, as well as the position of \psr and PSR~J1809$-$1917, a nearby pulsar enclosed by the TeV source HESS~J1809$-$193. In this analysis, the emission of the neighbouring source HESS~J1809$-$193 is removed by adding a two-component model, following the description derived in a recent analysis of this region \cite{J1809}.

We observe a compact bright emission centered around the pulsar, as well as a fainter, extended emission in the region around the pulsar. We account for the emission around \psr by adding a Gaussian source model with an extension of $\sigma = 0.056^\circ \pm 0.003^\circ$. The best-fit results derived for this source model agree within the error with the previously results for \source \cite{HGPS}. Hereafter, we will refer to this emission component as component A. After the removal of this component by modeling the emission, we observe a remaining large-scale $\gamma$-ray emission component. We find that this emission can be best described by an asymmetric Gaussian model, aligned along the galactic plane. We will refer to this source model as component B. The best-fit values for the two source models used to account for the emission around \psr, as well as the source models used to account for the emission in the vicinity of PSR~J1809$-$1917, are indicated by the black dashed lines in figure \ref{hess_sig}.

\section{\fermilat Data Analysis and Results}
In a previous analysis of the \fermilat data of the region around \psr using the standard LAT analysis software \texttt{Fermitools}, extended emission has been discovered \cite{fermi_araya}. This source, referred to as \fermi, is positionally coincident with \psr and \source, but with an extension of $0.6^\circ$, no connection could be made with the TeV emission. We reanalyse the data using \texttt{gammapy, version 0.18.2}, which enables us to optimise the morphology and spectrum of a source model simultaneously.

\begin{figure}
\centering
\includegraphics[width=0.8\textwidth]{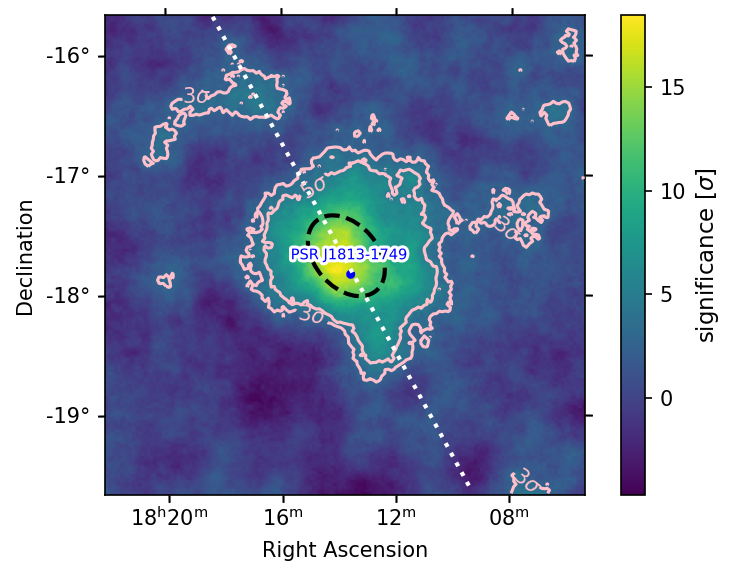}
\caption{Significance map of the region containing \fermi as seen by \fermilat. Additional emission in the field of view attributed to other astrophysical sources has been removed using their description in the 4FGL source catalogue. $3\,\sigma$ and $5\,\sigma$ contours of the residual emission in the field of view are overlaid in pink. The best-fit model is for \fermi is indicated by the black dashed line}
\label{fermi_sig}
\end{figure}

For this study, data from the beginning of the \fermilat mission in August 2008 until October 2021 is used. We use the most recent IRFs from Pass 8 version 3 --- \texttt{P8R3\_SOURCE\_V2}~\cite{pass8}, and the corresponding background models \texttt{iso\_P8R3\_SOURCE\_V3\_v1.txt}, \texttt{gll\_iem\_v07.fits}. Other sources in the $6^\circ$ Region-of-Interest (ROI) are accounted for by using source models from the 12-year 4FGL source catalogue~\cite{4FGL}. We allow a maximum zenith angle of $90^\circ$ for all events to avoid the inclusion of secondary $\gamma$-rays from the Earth's horizon, and use a bin size of $0.025^\circ$, as well as 8 energy bins per decade with logarithmic spacing. In order to avoid a large point-spread function we adopt an energy threshold of $1\,$GeV up to $1\,$TeV.

Figure \ref{fermi_sig} shows the significance map of the region around \psr, as seen by \fermilat. The emission from all sources in the ROI, except \fermi, has been accounted for by the respective source model in the 4FGL catalog. We again overlay the galactic plane as a white dotted line, the position of \psr, as well as $3\,\sigma$ and $5\,\sigma$ contours of the emission. Similarly to previous analyses of the region \cite{fermi_araya,fermi_xin}, we observe extended emission around the pulsar \psr, positionally coincident with \source. In contrast to the previous studies, using a disk model to describe the emission, we find that the emission can be described best by an asymmetric Gaussian model, with an extension of $0.38^\circ$, and an alignment along the galactic plane. The best-fit morphology of this model is indicated by the black dashed lines in figure \ref{fermi_sig}.

This emission is positionally coincident with the results derived in the \hess data for component B, though less extended. The addition of a second, compact component, coincident with component A in the \hess data, improves the description of the region, but cannot be established at a $5\,\sigma$ level and is therefore omitted for this study. 

\section{Energy-dependent behaviour of the emission}

On account of the positional coincidence between the detected $\gamma$-ray emission and the pulsar \psr, a leptonic origin of the emission is plausible. Previous studies of such systems have shown an energy-dependent morphology caused by electron diffusion and cooling \cite{enedep1,enedep2,J1825}. 

\vspace{0.08\textwidth}
\begin{minipage}{\textwidth}
  \hspace{-0.9cm}
  \begin{minipage}[b]{0.39\textwidth}
    \centering
    \begin{tabular}{c | c  }       
        \hline\hline  
        \noalign{\smallskip}
        Energy band & Energy [GeV]   \\    
        \noalign{\smallskip}
        \hline 
        \noalign{\smallskip}
        1 & $1.0\,-\,2.0$   \\
        2 & $2.0\,-\,4.0$    \\
        3 & $4.0\,-\,7.5$   \\
        4 & $7.5\,-\,18$  \\
        5 & $18\,-\,58$   \\
        6 & $58\,-\,1.0 \times 10^{3}$  \\
        \noalign{\smallskip}
        \cdashline{1-2}
        \noalign{\smallskip}
        H1 & $(0.4\,-\,1.3) \times 10^{3}$   \\      
        H2 & $(1.3\,-\,5.7) \times 10^{3}$   \\
        H3 & $(5.7\,-\,100) \times 10^{3}$  \\
        \noalign{\smallskip}
        \hline     
        \hline 
    \end{tabular}
    \captionof{table}{Energy bands used to test for energy-dependent morphology. Fermi bands are shown as 1- 6 and HESS bands as H1-H3}
    \end{minipage}
    \hspace{0.02\textwidth}%
    \begin{minipage}[b]{0.59\textwidth}
    \centering
    \includegraphics[scale=0.65]{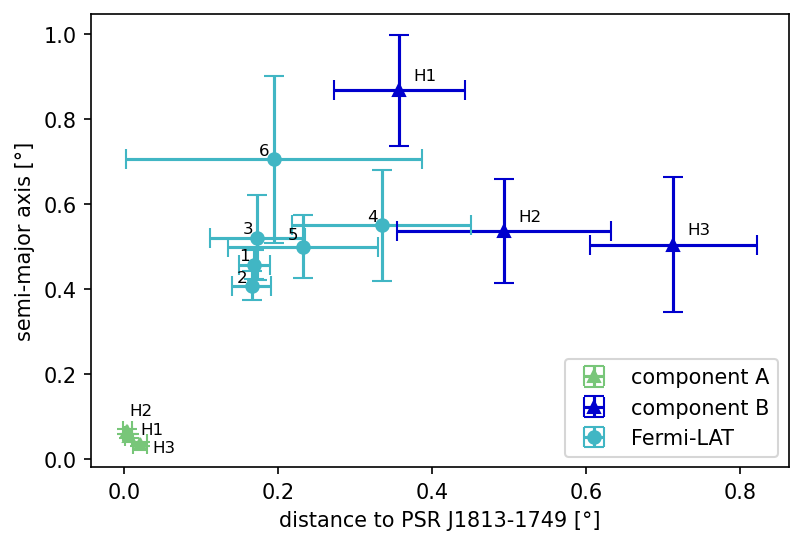}
    \label{enedep}
    \captionof{figure}{The distance between the centre of the best-fit model and the position of \psr is estimated in each energy bin and plotted against the extension of the semi-major axis.}
  \end{minipage}
  \end{minipage}
\vspace{0.08\textwidth}

To test for an energy-dependent morphology in the emission observed in this study the datasets are divided into six logarithmically spaced energy bins for the \fermilat data, as well as three energy bins for the \hess data. This binning was chosen based on the available statistics for each dataset. The best-fit models described above are then fitted in each energy bin respectively and the distance of the centre of the best-fit component to the position of \psr is computed and compared to the extension of the semi-major axis in the respective energy bin. These results are shown in Figure \ref{enedep}. 

Due to the low statistics in the \hess dataset, as well as the low statistics in the high energy regime of the \fermilat dataset, this study is subject to large uncertainties for the respective measurements. The derived extensions are compatible within the errors, we do not find a significant indication of an energy-dependent morphology. However, the extension in the last energy bin of the \fermilat data is compatible with the extension derived in the first energy bin of the \hess data, suggesting that the emission from \fermi can be connected to the extended emission observed in the \hess dataset.

\section{Joint-Modeling}
The discovery of extended emission in the \hess data enables a continuous description of the region around \psr, while the search for energy-dependent morphology showed that component B and the extended emission observed in the \fermilat data show a comparable extension. To investigate this result, we perform a joint analysis. To describe the emission observed by \fermilat and \hess we add a symmetric model, as well as an asymmetric model to both datasets. The likelihood minimisation is then performed on both datasets at the same time, while the IRFs for each dataset are taken into account. 

This minimisation returns a compact component, centred at a best-fit position of $\text{R.A.} = (273.40 \pm 0.003)^\circ$, $\text{Dec} = (-17.832 \pm 0.003)^\circ$ with an extension of $\sigma = (0.056 \pm 0.003)^\circ$. Based on the shape of the SED derived in the analysis of the respective datasets, we use a logarithmic parabola spectral model as spectral model. The estimated best-fit spectral index is $\Gamma = 2.05 \pm 0.03$, with a curvature $\beta = 0.06 \pm 0.001$ and a flux normalization at $1\,$TeV of $(3.16 \pm 0.13)\times 10^{-12}\,\text{cm}^{-2}\text{s}^{-1}\text{TeV}^{-1}$. 

For the second component, we find a best-fit position of $\text{R.A.} = (273.39 \pm 0.03)^\circ$, $\text{Dec} = (-17.504 \pm 0.04)^\circ$ with an extension of $\sigma = (0.54 \pm 0.03)^\circ$. This model is asymmetric, with an eccentricity of $e=0.73 \pm 0.04$ and a position angle of $\varphi = (213.76 \pm 7.22)^\circ$. We again used a logarithmic parabola model with a spectral index of $\Gamma = 2.17 \pm 0.03$, a curvature of $\beta = 0.04 \pm 0.001$, and a flux normalisation at $1\,$TeV of $(6.47 \pm 0.43 )\times 10^{-12}\,\text{cm}^{-2}\text{s}^{-1}\text{TeV}^{-1}$.

\begin{figure}
\centering
\includegraphics[width=0.8\textwidth]{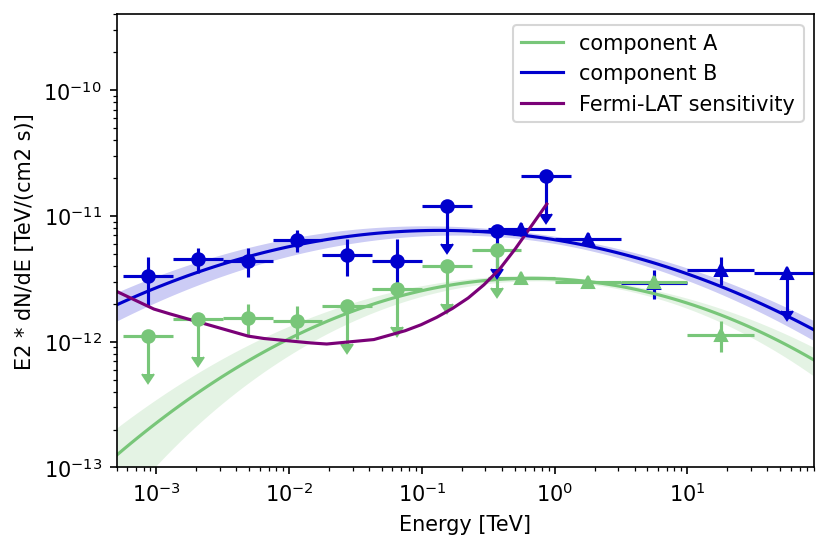}
\caption{Spectrum and SED for component A and component B derived from a joint-fit to both datasets. The broadband sensitivity of the \fermilat is indicated by the purple line \cite{sensitivity}.}
\label{spectrum}
\end{figure}

The spectrum and the spectral energy distribution (SED) for both components can be seen in Figure \ref{spectrum}. Additionally, the broadband sensitivity of \fermilat for sources located in the galactic plane and 12 years of data taking is overlaid \cite{sensitivity}. Most of the predicted emission from component A is below the detectable flux for the LAT, explaining why the emission could not be observed in the analysis of the \fermilat data. After removing the emission from components A and B using the best-fit parameters derived in this joint analysis, no emission remains in the \hess data. The addition of component A to the \fermilat dataset does not overestimate the emission in the region and the significance map is reasonably flat, indicating that this two-component source model is a good description of the region in the GeV and the TeV data.  

\section{Conclusion and Outlook}
In previous studies, the emission from \source and \fermi could not be connected, despite their positional coincidence, because of disagreements in the extension measurements. Due to these disagreements, all attempts to draw conclusions on the physical origin of the emission have used either the SED derived in the analysis of the GeV data, or the TeV data separately. As a consequence, the TeV source \source remains to be a source with an unknown origin. 

We reanalyse the region using an improved background model for the analysis of the TeV data, and increased exposure in the GeV data. We find that the emission in the \hess data can be well described by a two-component model. The first component, a compact Gaussian component with an extension of $0.06^\circ$, referred to as component A in this analysis, can be connected to \source. The detection of a second, extended component in this study makes it possible, for the first time, to connect the emission from \fermi to the emission observed in the TeV range and, therefore find a continuous description of the region around \psr over five decades of energy. Utilising these new results, we will be able to further investigate the physical origin of the emission. 

The observed emission is coincident with the pulsar \psr, but also with the SNR~G12.82$-$0.02 and the stellar cluster Cl~1813$-$178. Therefore the emission could be caused by electrons escaping the confines of the pulsar and forming a pulsar wind nebula, which we would then observe as \source. Some of these electrons further escape into the ISM, forming the halo-like structure observed as \fermi and component B. 
Another possibility is that the observed $\gamma$-ray emission is caused by protons from the cosmic ray sea. These protons are accelerated at the shock fronts of the SNR or the stellar cluster. producing $\gamma$-rays through the interaction with photons from molecular clouds in the region. A forthcoming publication will examine these possible emission scenarios in depth and further investigate possible evolution scenarios of the system around \psr.

\section{Acknowledgments}
This work is supported by the Deutsche Forschungsgemeinschaft (DFG, German Research Foundation) – Project Number 452934793.

\bibliographystyle{aa}
\bibliography{mybibliography}



\clearpage
\section*{Full Authors List: \hess\ Collaboration}
\scriptsize
\noindent
F.~Aharonian$^{1,2,3}$, 
F.~Ait~Benkhali$^{4}$, 
A.~Alkan$^{5}$, 
J.~Aschersleben$^{6}$, 
H.~Ashkar$^{7}$, 
M.~Backes$^{8,9}$, 
A.~Baktash$^{10}$, 
V.~Barbosa~Martins$^{11}$, 
A.~Barnacka$^{12}$, 
J.~Barnard$^{13}$
% ...

\medskip

\noindent
$^{1}$Dublin Institute for Advanced Studies, 31 Fitzwilliam Place, Dublin 2, Ireland\\
$^{2}$Max-Planck-Institut f\"ur Kernphysik, P.O. Box 103980, D 69029 Heidelberg, Germany\\
$^{3}$Yerevan State University,  1 Alek Manukyan St, Yerevan 0025, Armenia\\
$^{4}$Landessternwarte, Universit\"at Heidelberg, K\"onigstuhl, D 69117 Heidelberg, Germany\\
$^{5}$IRFU, CEA, Universit\'e Paris-Saclay, F-91191 Gif-sur-Yvette, France\\
$^{6}$Kapteyn Astronomical Institute, University of Groningen, Landleven 12, 9747 AD Groningen, The Netherlands\\
$^{7}$Laboratoire Leprince-Ringuet, École Polytechnique, CNRS, Institut Polytechnique de Paris, F-91128 Palaiseau, France\\
$^{8}$University of Namibia, Department of Physics, Private Bag 13301, Windhoek 10005, Namibia\\
$^{9}$Centre for Space Research, North-West University, Potchefstroom 2520, South Africa\\
$^{10}$Universit\"at Hamburg, Institut f\"ur Experimentalphysik, Luruper Chaussee 149, D 22761 Hamburg, Germany\\
$^{11}$Deutsches Elektronen-Synchrotron DESY, Platanenallee 6, 15738 Zeuthen, Germany\\
$^{12}$Obserwatorium Astronomiczne, Uniwersytet Jagiello{\'n}ski, ul. Orla 171, 30-244 Krak{\'o}w, Poland\\
$^{13}$Department of Physics, University of the Free State,  PO Box 339, Bloemfontein 9300, South Africa\\
% ...

%
%
%

\end{document}